\RequirePackage{fix-cm}
\documentclass{svjour3}                     
\smartqed  
\usepackage{graphicx}
\usepackage{amssymb}
\usepackage{color}
\usepackage{amsmath}
\usepackage{mathptmx}      
\newcommand{\jt}[1]{{\textcolor{black}{#1}}}
\begin{document}

\title{ Friedmann cosmology with decaying vacuum density in Brans-Dicke theory
}



\author{C. P. Singh        \and  Joan Sol\`a Peracaula
}


\institute{C. P. Singh  \at
              Department of Applied Mathematics,
 Delhi Technological University,
 Bawana Road, Delhi-110 042, India.\\
              \email{cpsphd@rediffmail.com}           
           \and
          Joan Sol\`a  Peracaula \at
Departament de F\'isica Qu\`antica i Astrof\'isica, and Institute of Cosmos Sciences, Universitat de Barcelona, \\ Av. Diagonal 647, E-08028 Barcelona, Catalonia, Spain\\
\email{sola@fqa.ub.edu}
}

\date{Received: date / Accepted: date}

\maketitle

\begin{abstract}
In this paper, we study Friedmann cosmology with time-varying vacuum energy density in the context of Brans-Dicke theory. We consider an isotropic and homogeneous flat space, filled with a matter-dominated perfect fluid and a dynamical cosmological term $\Lambda(t) $, obeying the equation of state of the vacuum. As the exact nature of a possible time-varying vacuum is yet to be found, we  explore $\Lambda(t)$ given by the phenomenological law $\Lambda(t)=\lambda+\sigma H$, where $\lambda$ and $\sigma$ are positive constants. \jt{We solve the model and then focus on  two different cases $\Lambda_{H1}$ and $\Lambda_{H2}$  by assuming $\Lambda=\lambda$ and $\Lambda=\sigma H$, respectively.  Notice that $\Lambda_{H1}$ is the analog of the standard $\Lambda$CDM, but within the Brans-Dicke cosmology.} We find the analytical solution of the main cosmological functions such as the Hubble parameter, the scale factor, deceleration and equation of state parameters for these models. \jt{In order to test the viability of the cosmological scenarios, we perform two sets of joint observational analyses of the recent Type Ia supernova data (Pantheon), observational measurements of Hubble parameter data, Baryon acoustic oscillation/Cosmic microwave background data and Local Hubble constant for each model. For the sake of comparison, the same data analysis is performed for the $\Lambda$CDM model. Each model shows a transition from decelerated phase to accelerated phase and can be viewed as an effective quintessence behavior.  \jt{Using the model selection criteria AIC and BIC to distinguish from existing dark energy models, we find that the Brans-Dicke analog of the $\Lambda$-cosmology (i.e.  our model $\Lambda_{H1}$)  performs at a level comparable to the standard $\Lambda$CDM, whereas  $\Lambda_{H2}$ is  less favoured.}}\\
\end{abstract}

\section{Introduction}
\label{intro}
The recent observational data from Type Ia supernova\cite{per1,adam,ast}, cosmic microwave background radiation \cite{sper}, galaxy clustering \cite{fel} and other cosmological observations \cite{kom,kom1,san,ade1,ade2} suggest that our Universe is currently experiencing a phase of accelerated expansion. It has been learnt that the Universe is dominated by dark energy (DE) with negative pressure which provides the dynamical mechanism for the accelerating expansion of the Universe. However, the nature of this substance is still undetermined. The cosmological constant (CC), initially introduced by Einstein, is a natural candidate of DE.  Such model is also known as standard Lambda-cold dark matter ($\Lambda$CDM) model. \jt{ In it $\Lambda=$const.  and sometimes it is referred to for short as  the $\Lambda$-cosmology}. In this scenario, DE is associated to the energy density of the quantum vacuum $\rho_{\Lambda}=\Lambda/8\pi G$. However, it faces a long-standing cosmological constant problem \cite{wee}. This CC problem stems from tremendous discrepancy between the theoretical value associated with quantum vacuum energy and the value required to confirm with observations.  \jt{As a matter of fact, all sorts of cosmological models predict a large value of the DE and they  require of an unnatural fine tuning to solve such discrepancy. In this sense the vacuum energy is not to be blamed more than many other DE models\cite{JSPRev2013}}.  Although several possible approaches have been adopted to explain or alleviate the CC problems \cite{cope}, there is no convincing fundamental theory for why vacuum energy dominance happened only recently \jt{and why its value is currently so close to the matter energy density (the so-called cosmic coincidence problem).}

\indent One possibility to \jt{mitigate certain aspects of the CC problem} is to consider  time evolving vacuum models, $\Lambda=\Lambda(t)$. A great deal of attention was dedicated to this possibility even before the discovering of the accelerating Universe \cite{ozer,peeb,car,lima1,over}.  $\Lambda(t)$ models may be an important alternative to the $\Lambda$CDM model.  \jt{The original proposals were essentially phenomenological\,\cite{over}, but a new generation of proposals are theoretically better rooted. } They are based on the idea that DE is the manifestation of vacuum quantum fluctuations in the curved space-time, after a renormalization in which the divergent vacuum contribution in the flat space-time is subtracted. The resulting effective vacuum energy density will depend on the space-time curvature, decaying from high initial values to smaller ones as the Universe expands. \jt{This idea underlies e.g. the class of  running vacuum models\,\cite{ShapSol,Fossil} (see \cite{JSPRev2013} for a review and references therein)  and it has been supported by recent calculations in quantum field theory  in curved spacetime\,\cite{Cristian2020}. }

\indent In recent years, a large class of flat non-singular Friedmann-Robertson-Walker type cosmologies, where the vacuum energy density evolves like a truncated power-series in the Hubble parameter H, have been discussed in the literature \cite{sch,carn,bor,ca,ca1,bas,bas1,per,den,lima2,mar,jay}. The functional form of $\Lambda(t)$ in most of them has usually been proposed on phenomenological grounds as it occurs with the vast majority of DE models. In this regard, a viable form of decaying vacuum energy density, namely $\Lambda(a)\propto H$,  was proposed by Sch\"{u}tzhold \cite{sch}.  \jt{ Such proposal was subsequently extended in the literature in the form of the the so-called ghost dark energy models\,\cite{GDE1,GDE2,GDE3}. These models together with the aforementioned class of running vacuum models\,\cite{JSPRev2013,JSPRev2015} both use  expansions of the vacuum energy density in powers of $H$, but of a different kind.  These examples show the significant interest raised by the dynamical dark energy models from different perspectives.}

\indent In the present paper, we focus our attention on the analytical and observational aspects of  the $\Lambda(t)$ models in the scalar tensor theory proposed by Brans and Dicke \cite{bd}. Brans-Dicke (BD) theory was the first gravity theory in which the dynamics of gravity were described by a scalar field while spacetime dynamics were represented by the metric tensor.  In this theory, the gravitational constant $G$ is replaced with a inverse of time-dependent scalar field $\phi$, which couples to gravity with a coupling parameter $\omega$ (Brans-Dicke parameter). This theory passed the experimental tests from the solar system \cite{ber}. In recent years, this theory got a new impetus as it arises naturally as the low energy limit of many theories of quantum gravity such as superstring theory or Kaluza-Klein theory. An attractive feature of BD theory is that the scalar field is a fundamental element of the theory, quite contrary to other models in which the scalar field is introduced separately in an ad hoc manner. The studies on Friedmann-Robertson-Walker model in the framework of BD theory have been carried out in Refs.\cite{pim,vp,sing1,ss,nb,nb1,sen,mo,das,ar,ari,xu,sing2,kar,sing3,sing4,sing5,sing6,ms,sing7}.\\

\indent The aim of the work is to find a cosmological scenario of the model in BD theory with varying cosmological term which would be capable to link the dynamics of the early Universe with that of our late Universe. This work extends the successful approach recently presented on BD cosmology with a rigid cosmological term \cite{BD1920,js1}, and  reinforces the idea that dynamical models of the vacuum energy can be very helpful to improve the fit to cosmological and cosmographical observations \cite{RVMfit,js2,js3,Rezaei1,Rezaei2}. In particular, they help alleviating the so-called  $H_0$ and $\sigma_8$ tensions\cite{Verde19,js4,Intertwined}, see e.g. \cite{RVMtensions1,RVMtensions2} and the very recent work \cite{EPLPersp}. In the present study, we compare  the consequent cosmological scenario with the constraints imposed by the observational data of Type Ia supernova (Pantheon), observational Hubble parameter data, baryon acoustic oscillations data/cosmic microwave background and local $H_0$. The analysis of Hubble- redshift relation has shown a good fit with best fit values of the model parameters.\\
\indent The structure of the paper is as follows. In Section \ref{section:2}, we introduce the basic cosmological equations. The solution of the field equations is presented in Section \ref{section:3} with time varying cosmological constant. Section \ref{section:4} describes and  places constraints on the main parameters of our vacuum models by performing two sets of joint likelihood analysis consisting of Type Ia supernova (Pantheon) data, the observational Hubble parameter data (OHD), baryonic acoustic oscillations/ cosmic microwave background (BAOs/CMB) data and local $H_0$. Section \ref{section:5} is divided in subsections. In Subsection 5.1, we discuss the evolution of the cosmological parameters using fitting values and in subsection 5.2, the model selection criterion is discussed.  Finally, in Section \ref{section:6}, we present the summary of the work.
\section{BD field equations with time-dependent vacuum}\label{section:2}
\noindent The action for BD theory extended to the cosmological constant (CC) in Jordan frame reads as follows \cite{uh,kim,we}.
\begin{equation}\label{c0}
S=\int{d^4x \sqrt{-g}\left[\frac{1}{16\pi}\left(\phi R-\frac{\omega}{\phi}\nabla_{\alpha}\phi\;\nabla^{\alpha}\phi\right)-\rho_{\Lambda} +\mathcal{L}_{m}\right]},
\end{equation}
where $\phi$ is the BD scalar field representing the inverse of the Newton constant, which is allowed to vary with space and time, $\omega$ is the dimensionless constant which is known as a coupling parameter, or Brans-Dicke parameter,  of the theory and $\mathcal{L}_{m}$ is the matter Lagrangian. It is to be noted that there is no potential for BD scalar field $\phi$ in the original BD theory, however, we admit the presence of CC term associated with vacuum energy density, $\rho_{\Lambda}$.\\
\indent Variation of this action with respect to the metric $g_{\mu\nu}$ and the BD scalar field $\phi$ yield the following field equations, respectively.
\begin{equation}\label{c2}
G_{\mu\nu}=R_{\mu\nu}-\frac{1}{2} g_{\mu\nu}R =\frac{8\pi}{\phi}\left(T^{m}_{\mu\nu}-g_{\mu\nu}\rho_{\Lambda}\right)+\jt{\frac{8\pi}{\phi}}T^{BD}_{\mu\nu}
\end{equation}
\noindent and
\begin{equation}\label{c3}
\Box\phi=\frac{8\pi}{(2\omega+3)}\left(T^{m\;\mu}_{\mu}-4\rho_{\Lambda}\right),
\end{equation}
where $T^{m}_{\mu\nu}$ is the the energy-momentum tensor of matter and $T^{m \;\mu}_{\mu}$ is the trace of $T^{m}_{\mu\nu}$, and other symbols have their usual meaning. It is convenient to introduce the effective energy-momentum tensor for the two fluids, matter and vacuum energy density through $\tilde{T_{\mu\nu}}=T_{\mu\nu}-g_{\mu\nu}\rho_{\Lambda} $ and adopts the perfect fluid form:
\begin{equation}\label{mc7}
\tilde{T}_{\mu \nu}=(\rho+p)u_{\mu}u_{\nu}+p g_{\mu\nu},
\end{equation}
where $\rho=\rho_m+\rho_{\Lambda}$ and $p=p_m+p_{\Lambda}$. We assume that the matter part contains the pressureless contribution of cold dark matter.  The vacuum energy density $\rho_{\Lambda}$ follows the usual equation of state (EoS) as $p_{\Lambda}=-\rho_{\Lambda}$. Also, $ T^{BD}_{\mu\nu}$ is the energy-momentum for the BD scalar which is defined by
\begin{equation}\label{tbd}
T^{BD}_{\mu\nu}= \frac{1}{8\pi}\Bigl[\frac{\omega}{\jt{\phi}}\left(\nabla_{\mu}\phi\nabla_{\nu}\phi-\frac{1}{2}g_{\mu\nu}\nabla_{\alpha}\phi
\nabla^{\alpha}\phi\right)+
\jt{\nabla_{\mu}\nabla_{\nu}\phi-g_{\mu\nu}\nabla_{\alpha}\nabla^{\alpha}\phi}\Bigr].
\end{equation}
\noindent Let us start with the homogeneous and isotropic flat Friedmann-Lema\^{\i}tre-Robertson-Walker (FLRW) line element
\begin{equation}\label{c1}
ds^2 = -dt^2 + a^2(t) \left[dr^2+r^2(d\theta^2+\sin^2\theta d\phi^2)\right],
\end{equation}
where $a(t)$ is the scale factor of the universe. Throughout we use units such that the speed of light, $c=1$.\\
\indent The field equations \eqref{c2} and \eqref{c3} for metric \eqref{c1} and energy-momentum tensors \eqref{mc7} and \eqref{tbd} are simplified to
\begin{equation}\label{c5}
3H^2+3H\frac{\dot{\phi}}{\phi}-\frac{\omega}{2}\frac{\dot{\phi^2}}{\phi^2}
=\frac{8\pi}{\phi}\rho,
\end{equation}
\begin{equation}\label{c4}
\jt{2\dot{H}+3H^2+\frac{\ddot{\phi}}{\phi}+2H\frac{\dot{\phi}}{\phi}+\frac{\omega}{2}\frac{\dot{\phi^2}}{\phi^2}=-\frac{8\pi}{\phi} p},
\end{equation}
\begin{equation}\label{c7}
\ddot{\phi}+3H\dot{\phi}=
\frac{8\pi}{(2\omega+3)}(\rho-3p).
\end{equation}
where an overdot denotes derivative with respect to cosmic time $t$ and $H=\dot{a}/a$ is the Hubble parameter. The first equation \eqref{c5} corresponds to the Friedmann equation and the second equation \eqref{c7} is the equation of motion of the BD scalar field.\\
\indent If we ignore the inhomogeneities arising from the (linear) field perturbations, the BD field can be treated as a perfect fluid $T^{BD}_{\mu\nu}=(\rho_{BD}+p_{BD})u_{\mu}u_{\nu}+p_{BD} g_{\mu\nu}$ with energy and pressure are respectively given by
\begin{equation}\label{cc81}
\jt{\rho_{BD}=\frac{1}{8\pi }\left[\frac{\omega}{2}\left(\frac{\dot{\phi}^2}{\phi}\right)-3H\dot{\phi}\right]},
\end{equation}
\begin{equation}\label{cc82}
\jt{p_{BD}=\frac{1}{8\pi }\left[\frac{\omega}{2}\left(\frac{\dot{\phi}^2}{\phi}\right)+2H \dot{\phi}+\ddot{\phi}\right]}.
\end{equation}
\indent Finally, the geometric Bianchi identity of $\nabla_{\nu}G^{\mu\nu}=0$ in Eq.\eqref{c2}, which plays a role of the consistency relation, leads to
\begin{equation}\label{bi}
\nabla_{\nu}\left(R^{\mu\nu}-\frac{1}{2}g^{\mu\nu}R\right)=0=\nabla_{\nu}\left(\frac{8\pi}{\phi}\tilde{T}^{\mu\nu}+\jt{\frac{8\pi }{\phi}} T^{\mu\nu}_{BD}\right).
\end{equation}
\indent One interesting thing about working in Jordan frame is that the conservation equation holds for matter and scalar
field separately, i.e., equations of motion of matter do not enter into the BD scalar field. It means that $\tilde{T}^{\mu\nu}$ obeys the usual conservation law, $\nabla_{\nu}\tilde{T}^{\mu\nu}=0$, which takes the form
\begin{equation}\label{c8}
\dot{\rho_{m}}+3(\rho_m+p_m)\frac{\dot{a}}{a}=-\dot{\rho_{\Lambda}}.
\end{equation}
\indent In this paper, we study the model dominated by pressureless dark matter $(p_m=0)$ in BD theory. It is to be noted that the equation of state of the vacuum energy density maintains the usual form $p_{\Lambda}(t)=-\rho_{\Lambda}(t)=-\phi \Lambda(t)/8\pi$ despite the fact that $\Lambda(t)$ evolves with time. Now, from \eqref{bi} and because of matter conservation, we are left with\footnote{\jt{It may be illustrative to point out here that  if one would define the energy density and pressure of the BD field in a different form, namely in such a way that they would represent the exact departure of the BD theory from GR, then the corresponding tensor associated to these new quantities  $ T^{\mu\nu}_{BD}$ would be locally and covariantly conserved as it does in the case of matter, namely  $ \nabla_\mu T^{\mu\nu}_{BD}=0$.  This alternative formulation has been used in \cite{BD1920,js1}.}   }
\begin{equation}
\left(\nabla_{\nu}\{\frac{8\pi}{\phi}\}\tilde{T}^{\mu\nu}+\nabla_{\nu}\{\frac{8\pi}{\jt{\phi}}T^{\mu\nu}_{BD}\}\right)=0,
\end{equation}
which finally gives
\begin{equation}\label{cc8}
\dot{\rho}_{BD}+3\frac{\dot{a}}{a}\left(\rho_{BD}+p_{BD}\right)=\left(\frac{\dot{\phi}}{\jt{\phi}}\right)\left(\rho+\jt{\rho_{BD}}\right)\,.
\end{equation}
\noindent \jt{ Equation \eqref{cc8} is indeed a consistency condition originating from the Bianchi identity $\nabla_{\nu}G^{\mu\nu}=0$. It can indeed be checked that the covariant conservation laws \eqref{c8} and \eqref{cc8} can also be obtained upon lengthy but straightforward computation by combining Eqs. \eqref{c5} - \eqref{c7}, which are identical to that of general relativity (GR). Although the calculation is more involved than in GR, the final result turns out to be the same. Thus, we shall use \eqref{c5} and \eqref{c8} to obtain the solution of the model and finally we use \eqref{cc8} to get the consistency condition using the fitting values of the model parameters obtained from observational data (to be discussed in Sect.4).}\\
\indent In the framework of BD cosmology the BD scalar field $\phi$ \jt{one usually searches for  power-law relation in terms of scale factor} \cite{pam,ban,sh,GRF2018,JavierJoan2018,Karimkhani}, namely\footnote{\jt{While there is no a priori reason to assume that a power-law solution is viable, let us notice that the fractional variation of the effective gravitational coupling $G=\frac{1}{\phi}$ in BD theories is given by $\frac{\dot{G}}{G}=-\frac{\dot{\phi}}{\phi}=-\epsilon H$. Therefore, for sufficiently small $|\epsilon|$, this is consistent with the bounds on the time variation of the gravitational coupling\,\cite{Uzan}. Other partial justifications can be checked a posteriori, as we shall see.} }
\begin{equation}\label{c9}
\phi=\phi_0\; a(t)^\epsilon,
\end{equation}
where $\phi_0$  and $\epsilon$ are constants. A case of particular interest is that when $\epsilon$ is small whereas $\omega$ is large so that the product $\omega\epsilon$ results of order unity \cite{ban,GRF2018,JavierJoan2018}. This is interesting because local experiments set a very high lower bound on $\omega$. This choice with small $\epsilon$ can lead to consistent results which may justify this specific choice among other possible choices \cite{ban}. The Cassini experiment \cite{ber} implies that $\omega > 10^4$. Likewise, as previously indicated, a slow fractional variation of $\phi$ will lead to a small fraction variation of $G$, consistent with observations. Therefore, it is clear that the interesting case is that one when $\epsilon$ is small whereas $\omega$ is large, so that the product $ \omega \epsilon$ results of order unity. As for $\omega$, it is usually assumed large, but we do not find $\omega$ so large  because we assume that the Cassini bound on this parameter only applies to the astrophysical domain, not to the cosmological one. This is admissible because of the possible existence of screening effects (chamaleon etc) which can operate in the local domain. These  effects do not apply at the cosmological level and permit a discussion of the BD framework free from the stringent Cassini bounds\,\cite{Avilez,Clifton}.   See e.g. \cite{js1}  for a detailed discussion.\\
\indent We also note that in Ref.\cite{GRF2018,JavierJoan2018} this kind of power-law solution is used to show that BD cosmology with a cosmological term can  mimic the running vacuum model, which is very convenient in order to improve the fitting of the cosmological data \cite{JavierJoan2018}. We expect that this feature will also help here. \\
\indent With the above assumption, the Eq.\eqref{c5} is rewritten as
\begin{equation}\label{c10}
H^2=\frac{2}{(6+6\epsilon-\omega \epsilon^2)}\frac{8\pi}{\phi}(\rho_m+\rho_{\Lambda}),
\end{equation}
where $H=\dot{a}/{a}$ is the Hubble parameter and $\Lambda=8\pi \rho_{\Lambda}/\phi$. It can be observed that in the limit of $\epsilon \rightarrow 0$, the standard cosmology is recovered. To make the Eq.\eqref{c10} to have physical meaning, i.e., to make $(6+6\epsilon-\omega \epsilon^2)>0$, one has the following constraint on the value of $\epsilon$ which is given by $\frac{3-\sqrt{9+6\omega}}{\omega}< \epsilon < \frac{3+\sqrt{9+6\omega}}{\omega}$, where $\omega >0$.\\
\indent Finally, combining equations \eqref{c8} and \eqref{c10}, we find
\begin{equation}\label{c11}
\dot{H}+\frac{(3+\epsilon)}{2}H^2=\frac{3\Lambda}{(6+6\epsilon-\omega \epsilon^2)}.
\end{equation}
In what follows, we investigate the cosmic evolution with a class of time evolving vacuum models. \jt{Notice that up to this point the above equations are valid for any $\Lambda$, not necessarily a constant, it can be a function of the cosmic time. Recall that the equation of state remains $p_{\Lambda}(t)=-\rho_{\Lambda}(t)$.  In the next sections, however, we specify some possible forms.}
\section{\jt{Brans-Dicke theory with time-varying $\Lambda$}}\label{section:3}
In this paper we parameterize the functional form of $\Lambda(t)$ as a combination of constant term and some multiple of the Hubble parameter, i.e.,
\begin{equation}\label{c12}
\Lambda(t)=\lambda+\sigma H,
\end{equation}
where $\lambda$ and $\sigma$ are positive constants. The model with $\lambda=0$, hence  $\Lambda \propto H$, was discussed in Refs.\cite{sch,bor,ca,ca1}, \jt{but only in the context of GR.  In \cite{GRF2018,JavierJoan2018} this case was studied from the point of view of its ability to emulate the running vacuum model. In the following subsections, we study the two extreme situations $(\lambda\neq 0, \sigma=0)$ and $(\lambda=0, \sigma\neq 0)$  in the BD framework and perform the corresponding observational analysis.  The detailed solution of the general class of models \eqref{c12} for any value of $\lambda$ and $\sigma$ is given in an  Appendix. }

\subsection{\jt{$\Lambda_{H1}-$ model: the standard $\Lambda$ cosmology in BD theory}}\label{subsection:3.1}
In this section, we consider the  $\Lambda=$ const. cosmology in the context of BD theory in order to appreciate the differences with respect to the $\Lambda(t)$ model explored subsequently. Assuming $\sigma=0$ in Eq. \eqref{c12}, we have $\Lambda(t)=\lambda=const.$ (hereafter $\Lambda_{H1}-$ model). Thus, the vacuum term in \eqref{c12} is constant and given by
\begin{equation}\label{c13}
\Lambda_0=\lambda=3\Omega_{\Lambda} H^2_0,
\end{equation}
\noindent where $\Lambda_0$, $H_0$ and $\Omega_{\Lambda}$ are the current value of vacuum energy density, Hubble parameter and density parameter at present epoch $t=t_0$, respectively. \\
\indent Using \eqref{c13}, the evolution equation \eqref{c11} reads
\begin{equation}\label{c0}
\frac{d\;h^2}{dx}+(3+\epsilon)h^2=\frac{18\;\Omega_{\Lambda}}{(6+6\epsilon-\omega \epsilon^2)},
\end{equation}
where $h=H/H_0$ is the dimensionless Hubble parameter and $x=\ln\;a$. Solving Eq.\eqref{c0}, we obtain the Hubble function in terms of redshift $z$ as
\begin{equation}\label{c14}
H(z)=H_0\left[\frac{18\Omega_{\Lambda}}{(6+6\epsilon-\omega \epsilon^2)(3+\epsilon)}+\left(1-\frac{18\Omega_{\Lambda}}{(6+6\epsilon-\omega \epsilon^2)(3+\epsilon)}\right)(1+z)^{(3+\epsilon)}\right]^{1/2},
\end{equation}
where $(1+z)=a_0/a$. We can check that $\epsilon=0$ reduces Eq. \eqref{c14} to the corresponding equation in $\Lambda$CDM model as expected. We can define the normalized Hubble expansion as a function of redshift
\begin{equation}\label{c15}
h^2(z)=\frac{H^2(z)}{H^2_0}=\tilde{\Omega_{\Lambda1}}+\tilde{\Omega_{m1}}(1+z)^{(3+\epsilon)},
\end{equation}
where we have used the following parametrization
\begin{equation}\label{c16}
\tilde{\Omega_{\Lambda1}}=1-\tilde{\Omega_{m1}}=\frac{18\;\Omega_{\Lambda}}{(6+6\epsilon-\omega \epsilon^2)(3+\epsilon)}
\end{equation}
As expected, for $\epsilon\rightarrow 0$, we have $\tilde{\Omega_{\Lambda1}}\sim \Omega_{\Lambda}$. Thus, the traditional cosmology is a particular solution of the $\Lambda_{H1}$ model with $\epsilon$ strictly equal to zero.\\
\indent The scale factor of the Universe, normalized to unity at the present epoch, is given by
\begin{equation}\label{c17}
a_{\Lambda}(t)=\left(\frac{\tilde{\Omega_{m1}}}{\tilde{\Omega_{\Lambda1}}}\right)^{1/(3+\epsilon)}\;
\left[\sinh\left(\frac{(3+\epsilon)\sqrt{\tilde{\Omega_{\Lambda1}}}}{2}\;H_0 t\right)\right]^{2/(3+\epsilon)}.
\end{equation}
The Hubble parameter in terms of cosmic time $t$ is given by
\begin{equation}\label{c18}
H(t)=H_0\sqrt{\tilde{\Omega_{\Lambda1}}}\;\coth\left(\frac{(3+\epsilon)\sqrt{\tilde{\Omega_{\Lambda1}}}}{2}\;H_0 t\right).
\end{equation}
The cosmic time is related with the scale factor as
\begin{equation}\label{c19}
t_{\Lambda}(a)=\frac{2}{(3+\epsilon)\sqrt{\tilde{\Omega_{\Lambda1}}}H_0}\;\sinh^{-1}\left(\sqrt{\frac{\tilde{\Omega_{\Lambda1}}}{\tilde{\Omega_{m1}}}
}\;a^{(3+\epsilon)/2}\right)
\end{equation}
The current age of the Universe is given by
\begin{equation}\label{c20}
t_{0\Lambda}=\frac{2}{(3+\epsilon)\sqrt{\tilde{\Omega_{\Lambda1}}}H_0}\;\sinh^{-1}\left(\sqrt{\frac{\tilde{\Omega_{\Lambda1}}}{\tilde{\Omega_{m1}}}
}\right).
\end{equation}
From \eqref{c14}, one can  deduce the deceleration parameter $q$, which is defined as $q=-a\ddot{a}/\dot{a}^{2}$. It is given by
\begin{equation}\label{c21}
q(z)=-1+\frac{\frac{(3+\epsilon)}{2}\tilde{\Omega_{m1}}\;(1+z)^{(3+\epsilon)}}{\tilde{\Omega_{\Lambda1}}+\tilde{\Omega_{m1}}(1+z)^{(3+\epsilon)}}.
\end{equation}
The value of $q$ at present time ($z=0$) is given by
\begin{equation}\label{c22}
q(z=0)=-1+\left(\frac{3+\epsilon}{2}\right)\tilde{\Omega_{m1}}.
\end{equation}
The transition from deceleration to acceleration takes place for $\tilde{\Omega_{m1}}=2/(3+\epsilon)$. For any value $\tilde{\Omega_{m1}}<2/(3+\epsilon)$, the present-day cosmic expansion is accelerating. Now, it is also possible to find the transition redshift $z_{tr}$ at which the Universe transits from deceleration to acceleration, i.e.,
\begin{equation}\label{c23}
z_{tr}=\left[\frac{2}{(1+\epsilon)}\frac{\tilde{\Omega_{\Lambda1}}}{\tilde{\Omega_{m1}}}\right]^{1/(3+\epsilon)}-1
\end{equation}
\jt{For $\epsilon=0$ we can see that we recover the corresponding value in the concordance $\Lambda$-cosmology with GR,  as expected.}
The transition time is calculated as
\begin{equation}\label{c24}
t_{tr}=\frac{2}{(3+\epsilon)\sqrt{\tilde{\Omega_{\Lambda1}}}H_0}\sinh^{-1}\left(\sqrt{\frac{1+\epsilon}{2}}\right).
\end{equation}
\noindent Let us calculate the effective equation of state (EoS) parameter $w_{eff}$ \jt{for the compound fluid of the model}. An accelerated expansion of the Universe is possible only if the effective EoS parameter $w_{eff}$  satisfies $3w_{eff}+1<0$. The effective EoS parameter can be obtained by
\begin{equation}\label{ef}
w_{eff}=-1-\frac{1}{3}\frac{d(\ln h^2)}{dx},
\end{equation}
\noindent where $x=ln\;a$. Using \eqref{c15}, the EoS parameter \eqref{ef} is calculated as
\begin{equation}\label{c25}
w_{eff}=-1+\frac{(3+\epsilon)\tilde{\Omega_{m1}}}{3h^2}(1+z)^{(3+\epsilon)}.
\end{equation}
\jt{The $\epsilon$-dependent part represents  the departure from the GR result.  It is easy to check that the above EoS can be written as follows:}
\begin{equation}\label{c25b}
\jt{w_{eff}=-1+\frac{3+\epsilon}{3} \frac{1}{1+r a^{3+\epsilon}}\,,}
\end{equation}
\jt{where we have defined the ratio $r=\tilde{\Omega_{\Lambda1}}/\tilde{\Omega_{m1}}$.}
\jt{One can observe that $w_{eff}\rightarrow -1$ in the late time ($a\gg1$), whereas $ w_{eff}\rightarrow \frac{\epsilon}{3}$  in the remote past ($a\to 0$). Therefore, the model corresponds to de Sitter in future time and performs a transition  from a situation of  essentially matter dominance ($w_{eff}\simeq 0$) into a future one of vacuum dominance ($w_{eff} = -1$)}. The EoS does not cross the phantom divide line $w\leq -1$ which shows that the $\Lambda_{H2}$ model is free from big-rip singularity. The present value of EoS parameter is obtained by setting $a=1$ ($z=0$) in the above equation:
\begin{equation}\label{c26}
w_{eff}(z=0)=-1+\frac{(3+\epsilon)\tilde{\Omega_{m1}}}{3},
\end{equation}
where $h(z)=1$ at $z=0$. Therefore, the condition for $3w_{eff}(z=0)+1<0$ implies that $\tilde{\Omega_{m1}}<2/(3+\epsilon)$.\\
\indent Let us check the consistency condition \eqref{cc8} for the solution of this model. Using \eqref{cc81}, \eqref{cc82} and \eqref{c10}, Eq.\eqref{cc8} can be rewritten as\footnote{It is important to keep in mind that the exact consistency condition is actually $\epsilon$ times Eq.\,\eqref{lam1}.  By dividing out the exact equation by $\epsilon$ we are assuming that $\epsilon\neq 0$, as in fact it is the case in our fitting result (see Table 1).  However,  the presence of the additional factor of $\epsilon$ shows that Eq.\,\eqref{lam1} has a smooth limit to GR (for which $\epsilon=0$ exactly) and hence no such equation remains in that limit. Recall that the primary origin of the consistency condition is Eq.\,\eqref{cc8}, which disappears of course when there is no BD fluid \eqref{cc81}-\eqref{cc82} .} 
\begin{equation}\label{lam1}
2(\omega \epsilon-3)\dot{H}+ (\omega \epsilon^2+6\omega \epsilon-12)H^2=0
\end{equation}
Now, using the solution of $H$ obtained in Eq.\eqref{c14} into \eqref{lam1}, we obtain
\begin{equation}\label{lam2}
(\omega \epsilon-3)(3+\epsilon)(1-\tilde{\Omega_{\Lambda1}})a^{-(3+\epsilon)}-(\omega \epsilon^2+6\omega \epsilon-12)[\tilde{\Omega_{\Lambda1}}+(1-\tilde{\Omega_{\Lambda1}})a^{-(3+\epsilon)}]=0.
\end{equation}
The above equation gives a  relation between the constants for $a=a_0=1$, i.e., at present, which is given by
\begin{equation}\label{consistency1}
(\omega \epsilon-3)(3+\epsilon)(1-\tilde{\Omega_{\Lambda1}})-(\omega \epsilon^2+6\omega \epsilon-12)=0.
\end{equation}
\indent It is to be noted that one can use this equation for consistency  checkup and not for constraining  the parameters. In Section 5, we will present a detail discussion of the solutions obtained in this section by using best fit values.  \jt{We will check explictly the above consistency condition.}
\subsection{\jt{$\Lambda_{H2}-$ model: $\Lambda=\sigma H$ in BD theory}}\label{susection:3.2}
\noindent In this Section, we assume that the vacuum term is proportional to the Hubble parameter. This kind of cosmological model is a particular case of Eq.\eqref{c12} by setting $\lambda=0$ which is given by (hereafter, $\Lambda_{H2}$ model)
\begin{equation}\label{hh}
\Lambda(t)=\sigma H
\end{equation}
In Refs. \cite{sch,bor,ca,ca1}, the authors have studied the FLRW model with Eq.\eqref{hh} to describe the late time evolution of the Universe in the context of GR. In this paper, our aim is to study the FLRW model with this ansatz in a more dynamical framework of scalar -tensor theory as described by the Brans and Dicke theory.\\
\indent Utilizing this ansatz at the present epoch $\Lambda_0=\sigma H_0$ and taking into account that the current value of the vacuum energy density is $\Lambda_0=3\Omega_{\Lambda}H^2_{0}$, the parameter $\sigma$ is obtained to be
\begin{equation}\label{c27}
\sigma=3\Omega_{\Lambda}H_{0}.
\end{equation}
\indent Using \eqref{hh}, the evolution equation \eqref{c11} reduces to
\begin{equation}\label{c28}
\frac{dh}{dx}+\frac{(3+\epsilon)}{2}\;h=\frac{3\sigma}{(6+6\epsilon-\omega \epsilon^2)H_0},
\end{equation}
The solution of \eqref{c28} is given by
\begin{equation}\label{c29}
h=\frac{6\sigma}{(6+6\epsilon-\omega \epsilon^2)(3+\epsilon)H_0}+\left(1-\frac{6\sigma}{(6+6\epsilon-\omega \epsilon^2)(3+\epsilon)H_0}\right)(1+z)^{\frac{(3+\epsilon)}{2}}.
\end{equation}
Using \eqref{c27} into  \eqref{c29}, the Hubble parameter in terms of redshift can be given by
\begin{equation}\label{c30}
H=H_0\left[\frac{18\Omega_{\Lambda}}{(6+6\epsilon-\omega \epsilon^2)(3+\epsilon)}+\left(1-\frac{18\Omega_{\Lambda}}{(6+6\epsilon-\omega \epsilon^2)(3+\epsilon)}\right)(1+z)^{\frac{(3+\epsilon)}{2}}\right],
\end{equation}
We can define the normalized Hubble expansion as a function of redshift
\begin{equation}\label{c31}
h=\frac{H(z)}{H_0}=\tilde{\Omega_{\Lambda2}}+\tilde{\Omega_{m2}}(1+z)^{\frac{(3+\epsilon)}{2}},
\end{equation}
where we have used the following parametrization
\begin{equation}\label{c32}
\tilde{\Omega_{\Lambda2}}=1-\tilde{\Omega_{m2}}=\frac{18\;\Omega_{\Lambda}}{(6+6\epsilon-\omega \epsilon^2)(3+\epsilon)}.
\end{equation}
Thus, the Hubble rate of $\Lambda_{H2}$ is very different from Eq. \eqref{c15} of $\Lambda_{H1}$ model, which could have the different behavior when we do the observation. It is to be noted that in the absence of BD theory ($\epsilon =0$), Eq. \eqref{c31} reduces to the Eq. (70) of Ref.\cite{bas1}. The scale factor in normalized unit is given by
\begin{equation}\label{c33}
a(t)= \left[\frac{e^{\frac{(3+\epsilon)}{2}\tilde{\Omega_{\Lambda2}}H_0\;t}
-1+\tilde{\Omega_{\Lambda2}}}{\tilde{\Omega_{\Lambda2}}}\right]^{2/(3+\epsilon)}.
\end{equation}
From Eq.\eqref{c33}, it is observed that for small times (small compared to the present time), it can be approximated by
\begin{equation}
 a(t)\sim \left(1+\frac{(3+\epsilon)}{2}H_0\;t\right)^{2/(3+\epsilon)},
 \end{equation}
which has the same time dependence as in the standard flat BD model with dust. Therefore, the $\Lambda_{H2}$ model expands with decelerated rate in early time. The model predicts the Big-Bang in the past at the cosmic time: $t_{BB}=\frac{2H^{-1}_{0}}{(3+\epsilon)\;\tilde{\Omega_{\Lambda2}}}\;\ln(1-\tilde{\Omega_{\Lambda2}})$. The varying $\Lambda(t)$ starts dominating just at present time. In the limit of large times, that is, $(3+\epsilon)\tilde{\Omega_{\Lambda2}}H_0\;t\gg 1$ and $a\rightarrow \infty$, Eq.\eqref{c33} leads to $\exp(\tilde{\Omega_{\Lambda2}}H_0\;t)$, that is, the model tends to a de Sitter Universe.\\
\indent The Hubble parameter in terms of cosmic time $t$ is given by
\begin{equation}
H(t)=\frac{H_0\;\tilde{\Omega_{\Lambda2}}\;e^{\frac{(3+\epsilon)}{2}\tilde{\Omega_{\Lambda2}}H_0\;t}}{e^{\frac{(3+\epsilon)}{2}\tilde{\Omega_{\Lambda2}}H_0\;t}
-1+\tilde{\Omega_{\Lambda2}}}.
\end{equation}
In this model, the cosmic time is related to the scale factor as
\begin{equation}\label{c34}
t(a)=\frac{2H^{-1}_{0}}{(3+\epsilon)\;\tilde{\Omega_{\Lambda2}}}\;\ln[1+\tilde{\Omega_{\Lambda2}}(a^{2/(3+\epsilon)}-1)].
\end{equation}
It is straightforward to calculate the deceleration parameter in terms of redshift which takes the following form:
\begin{equation}\label{c35}
q=-1+\frac{(3+\epsilon)\tilde{\Omega_{m2}}\;(1+z)^{(3+\epsilon)/2}}{2\left[\tilde{\Omega_{\Lambda2}}
+\tilde{\Omega_{m2}}(1+z)^{(3+\epsilon)/2}\right]}.
\end{equation}
Note that for $z=0$, one finds the current value of deceleration parameter
\begin{equation}\label{c36}
 q_0=-1+\frac{(3+\epsilon)\tilde{\Omega_{m2}}}{2}.
 \end{equation}
Now, it is easy to check that the transition redshift, defined to be the zero point of the deceleration parameter, is given by
\begin{equation}\label{c37}
z_{tr}=\left(\frac{2\tilde{\Omega_{\Lambda2}}}{(1+\epsilon)\tilde{\Omega_{m2}}}\right)^{2/(3+\epsilon)}-1.
\end{equation}
\jt{For $\epsilon=0$ we can see that on this occasion we do not recover the corresponding value in the $\Lambda$-cosmology with GR. Compare, in contrast,  with Eq.\eqref{c23} of the $\Lambda_{H1}-$ model. This is because model $\Lambda_{H2}$ does not have a smooth connection with the concordance $\Lambda$CDM owing to the absence of a rigid cosmological term in it.}

The inflection point takes place at
\begin{equation}\label{c38}
t_{tr}=\frac{2H^{-1}_{0}}{(3+\epsilon)\tilde{\Omega_{\Lambda2}}}\;\ln\;\left(\frac{(3+\epsilon)\tilde{\Omega_{m2}}}{2}\right).
\end{equation}
The effective EoS for this model is given by
\begin{equation}\label{c39}
w_{eff}=-1+\frac{(3+\epsilon)\tilde{\Omega_{m2}}}{3h}\;(1+z)^{(3+\epsilon)/2}.
\end{equation}
We can observe that $w_{eff}\rightarrow -1$ in the late time. Therefore, the model corresponds to de Sitter in future time. The EoS does not cross the phantom divide line $w\leq -1$ which shows that the $\Lambda_{H2}$ model is free from big-rip singularity. The present-day value of $w_{eff}$ is obtained as
\begin{equation}\label{c40}
w_{eff}=-1+\frac{(3+\epsilon)\tilde{\Omega_{m2}}}{3}
\end{equation}
From Eq. \eqref{c40}, we can observe that the condition $3w_{eff}+1<0$  gives $\tilde{\Omega_{m2}} < 2/(3+\epsilon)$.  In particular, if $\epsilon=0$, we recover the expression of all the cosmological parameters obtained in Ref.\cite{ca}, where the model was treated within GR.\\
\indent Now, using the solution of $H$ obtained in Eq.\eqref{c30} into consistent eq.\eqref{lam1}, we obtain
\begin{equation}\label{lam3}
(\omega \epsilon-3)(3+\epsilon)(1-\tilde{\Omega_{\Lambda2}})a^{-\frac{(3+\epsilon)}{2}}-(\omega \epsilon^2+6\omega \epsilon-12)[\tilde{\Omega_{\Lambda2}}+(1-\tilde{\Omega_{\Lambda2}})a^{-\frac{(3+\epsilon)}{2}}]=0.
\end{equation}
The above equation gives a  relation between the constants for $a=a_0=1$, i.e., at present, which reads
\begin{equation}\label{consistency2}
(\omega \epsilon-3)(3+\epsilon)(1-\tilde{\Omega_{\Lambda2}})-(\omega \epsilon^2+6\omega \epsilon-12)=0.
\end{equation}
Again, in this model, one can use this equation for consistency checkup and not for constraining the parameters.

\section{Parameter estimation}\label{section:4}
In this section, we present the cosmic observations on the free parameters of $\Lambda_{H1}$ and $\Lambda_{H2}$ models. To this end we will use two joint observational set of data, as described below. We perform the goodness-of-fit of the models using Markov Chain Monte Carlo (MCMC) method by employing EMCEE python package \cite{emcee}.  We also perform the model selection criteria to determine favoured model. In what follows, we discuss the observational data which are to be used to constraint the parameters of the models.
\subsection{Hubble data}\label{subsection:4.1}
\noindent \jt{We use the Hubble data comprising of 36 measurements which includes 31 measurements from cosmic chronometric (CC) method \cite{moresco2012}, three correlated measurements from BAO signal in galaxy distribution \cite{alam2017}, and lastly two measurements from BAO signal in Ly-$\alpha$ forest distribution alone or cross-correlated with quasistellar objects (QSOs) \cite{delubac2015,font2014}.\\
\indent Thus, the chi-square function corresponding to 33 measurements of CC and Ly-$\alpha$ is defined as
\begin{equation}
\chi^2_{CC+Ly\alpha}=\sum_{i=1}^{33} \frac{[H_{obs}(z_i)-H_{th}(z_i)]^2}{\sigma_i^2},
\end{equation}
where $H_{th}(z_i)$ and $H_{obs}(z_i)$ represents theoretical and observed values, respectively, and $\sigma^{2}_{i}$ is the standard deviation of each $H_{obs}(z_i)$ as given in Table 2 of Ref.\cite{sivani2019}.\\
\indent Further, the chi-squared corresponds to the $3$ galaxy distribution measurements is given by
\begin{equation}
\chi^2_{gal}=A^T C^{-1} A
\end{equation}
where $C$ is the covariance matrix given by \cite{alam2017}
\begin{equation*}
	C=
	\begin{bmatrix}
		$3.65$ &$1.78$& $0.93$\\
		$1.78$& $3.65$ &$2.20$\\
		$0.93$& $2.20$& $4.45$\\
	\end{bmatrix}
\end{equation*}
and
\begin{equation*}
	A=
	\begin{bmatrix}
		H_{obs}(0.38)-H_{th}(0.38)\\
		H_{obs}(0.51)-H_{th}(0.51)\\
		H_{obs}(0.61)-H_{th}(0.61)\\
	\end{bmatrix}
\end{equation*}
Thus, the combined $\chi^2$ function for Hubble data is given by
\begin{equation}\label{ohd}
\chi^2_{H}=\chi^2_{CC+Ly\alpha}+\chi^2_{gal}.
\end{equation}
\subsection{Type Ia supernovae (Pantheon data)}
We use the Pantheon sample, the latest compilation of Type Ia supernovae (SNe) comprising  of $40$ binned data points in the redshift region $z \in [0.014,1.62]$ \cite{scolnic}.\\
\indent The $\chi^2$ function of the Pantheon SNe data is given by
\begin{equation}\label{snia}
\chi^2_{SNe(Pan)}= \Delta \mu^{T} \cdot C^{-1}\cdot \Delta \mu
\end{equation}
in which $\Delta\mu=\mu^{obs}_{i}-\mu^{th}$, where $\mu^{obs}_{i}$ is the observed distance modulus defined in Ref.\cite{beto} and  $\mu^{th}$, the theoretical distance modulus that depends on redshift and the cosmological parameters, is given by
 \begin{equation}
\mu^{th}=5\;\log_{10}[d_L(z)/10\; pc]+\mathcal{M},
\end{equation}
where $\mathcal{M}$ is the nuisance parameter. The quantity $d_{L}$, known as the dimensionless luminosity distance, is given by \cite{scolnic}
\begin{equation}
d_L(z)=(1+z)c \int_0^{z} \frac{ dz'}{H(z',\theta)},
\end{equation}
where $\theta$ represents the set of model parameters and $c$ is the speed of light.\\
\indent It should be noted that the covariance matrix  $C$ in \eqref{snia} is the sum of the systematic covariance $C_{sys}$ and statistical matrix $D_{stat}$ having a diagonal component \cite{scolnic,conley}.}
\subsection{BAO/CMB data set}
\noindent We use the combined baryon acoustic oscillation and cosmic microwave background (BAO/CMB) data from different observational missions \cite{santos}. We have taken the sample of BAO distances measurements from SDSS(R) \cite{padma2}, the 6dF Galaxy survey \cite{beut}, BOSS CMASS \cite{ander} and three parallel measurements from WiggleZ survey \cite{blake}. We combine theses results with the Planck 2015 \cite{ade}.\\
\indent We use measurements derived from the product of the CMB acoustic scale, and from the ratio of the BAO dilation scale to the sound horizon scale at the drag epoch. Thus, we can write the $\chi^2$ function as \cite{santos}
\begin{equation}
\chi_{BAO/CMB}^2=A^T C^{-1} A
\end{equation}
where $C^{-1}$ is the inverse of the covariance matrix \cite{santos} and $A$ is the matrix
\begin{equation*}
A=
\begin{bmatrix}
\frac{d_A(z_*,\theta)}{D_v(0.106,\theta)}-30.84\\
\frac{d_A(z_*,\theta)}{D_v(0.35,\theta)}-10.33\\
\frac{d_A(z_*,\theta)}{D_v(0.57,\theta)}-6.72\\
\frac{d_A(z_*,\theta)}{D_v(0.44,\theta)}-8.41\\
\frac{d_A(z_*,\theta)}{D_v(0.6,\theta)}-6.66\\
\frac{d_A(z_*,\theta)}{D_v(0.73,\theta)}-5.43
\end{bmatrix}
\end{equation*}
\noindent Here, $D_v(z,\theta)$ represents the dilation scale which is given by $D_v(z,\theta)=\left( \frac{d_A^2(z,\theta)\;zc}{H(z,\theta)}\right)^{1/3}$. The comoving angular diameter, $d_A(z,\theta)$ is defined as
\begin{equation}
d_A(z_*,\theta)=\int_{0}^{z_*}\frac{dz'}{H(z',\theta)}
\end{equation}
where $z_*$ indicates the photon  decoupling redshift and hold the value $z_*=1090$ as per the Planck 2015 results \cite{ade}. We have taken the correlation coefficient from Ref. \cite{hing}.
\subsection{Local Hubble constant}
\jt {We use $H_0 = 73.5 \pm 1.4$ $km \;s^{-1} Mpc^{-1}$ which is locally measured by SH0ES as reported in \cite{reid} in our analysis.}\\
\indent In order to constrain the model parameters with the above data sets we perform a Bayesian  Markov Chain Monte Carlo (MCMC) method. This method is based on the publicly available EMCEE package \cite{emcee} for analysing and plotting the contours. \jt{In our calculation, we have minimized the chi-square for two combinations of data set, which we believe are helpful for better fit values. The first one is labeled $DS1$ and contains $SNe(Pan)+H(z)+BAO/CMB+H_0$. The chi-square function for DS1 reads as  $\chi^2_{DS1}=\chi^2_{SNe(Pan)}+\chi^2_{H}+\chi^2_{BAO/CMB}+H_0$. The second is  $DS2$ which contains $SNe(Pan)+H(z)+BAO/CMB$ and the total chi-square function reads as $\chi^2_{DS2}=\chi^2_{SNe(Pan)}+\chi^2_{H}+\chi^2_{BAO/CMB}$.}
\begin{table}[h]
	\caption{The fitting values for the considered models ($\Lambda_{H1}$, $\Lambda_{H2}$ and $\Lambda$CDM) obtained from two different joint analysis of data sets, namely DS1:$SNe(Pan)+H(z)+ BAO/CMB+H_0$ and DS2:$SNe(Pan)+H(z)+ BAO/CMB$. }
	\centering
	\begin{tabular}{c | c c c | c c c  }
		\hline \hline
$ $ &  $ $ & DS1 & $ $ &   $ $ & DS2 & $ $   \\ [0.5ex]
\hline
$Parameter $ & $\Lambda_{H1}$ & $\Lambda_{H2}$ & $\Lambda CDM$ & $\Lambda_{H1}$ & $\Lambda_{H2}$ & $\Lambda CDM$ \\
\hline\\
$H_0$ & $71.090^{+0.743}_{-0.627}$ & $70.858^{+0.729}_{-0.969}$ & $71.545^{+1.175}_{-0.820}$ & $69.603^{+1.081}_{-1.100}$ & $68.643^{+0.942}_{-1.111}$ & $68.545^{+2.102}_{-1.742}$ \\\\
$\epsilon$ & $0.070^{+0.005}_{-0.005}$ & $0.406^{+0.053}_{-0.055}$ & $-$ & $0.068^{+0.005}_{-0.004}$ & $0.344^{+0.051}_{-0.044}$ & $-$ \\\\
$\omega$ &  $25.40^{+0.004}_{-0.003}$  & $15.783^{+7.744}_{-7.968}$  & $-$  & $26.231^{+0.002}_{-0.003}$  & $15.801^{+8.492}_{-10.684}$  & $-$  \\\\
$\Omega_{\Lambda}$ &  $0.73^{+0.133}_{-0.183}$  &  $0.74^{+0.102}_{-0.229}$  & $0.68^{+0.015}_{-0.010}$  & $0.72^{+0.142}_{-0.169}$  & $0.76^{+0.148}_{-0.123}$  &   $0.69^{+0.032}_{-0.028}$ \\

	\hline
	\end{tabular}
	
\end{table}

\begin{table}
	\caption{The numerical values of $a_{tr}$, $z_{tr}$, $q_0$, $w_{eff}(z=0)$ and $t_0$ using  best-fit results of model parameters }
	\centering
	\begin{tabular}{c | c c c | c c c}
		\hline \hline
$ $ &  $ $ & DS1 & $ $ &   $ $ & DS2 & $ $   \\ [0.5ex]
\hline
$Values$ & $\Lambda_{H1}$ & $\Lambda_{H2}$ & $\Lambda$CDM & $\Lambda_{H1}$ & $\Lambda_{H2}$ & $\Lambda$CDM\\
\hline\\
$z_{tr}$ & $0.574^{+0.360}_{-0.324}$ & $0.821^{+0.201}_{-0.103}$ &$0.701^{+0.024}_{-0.020}$ & $0.601^{+0.339}_{-0.343}$ & $0.763^{+0.245}_{-0.167}$ & $0.672^{+0.028}_{-0.025}$\\\\
$q_0$ & $-0.580^{+0.210}_{-0.160}$ & $-0.421^{+0.386}_{-0.288}$ & $-0.594^{+0.014}_{-0.018}$ & $-0.610^{+0.170}_{-0.190}$ & $-0.388^{+0.294}_{-0.630}$ & $-0.554^{+0.024}_{-0.030}$ \\\\
$w_{eff}(z=0)$ &  $-0.675^{+0.173}_{-0.123}$  & $-0.615^{+0.257}_{-0.259}$ & $-0.729^{+0.009}_{-0.012}$ & $-0.680^{+0.172}_{-0.120}$  & $-0.592^{+0.196}_{-0.435}$ & $-0.703^{+0.016}_{-0.020}$  \\\\
$t_0$ (in Gyrs) &  $13.69^{+1.829}_{-1.548}$  &  $14.18^{+1.419}_{-1.421}$  & $13.48^{+0.450}_{-0.230}$& $13.73^{+1.941}_{-1.419}$  & $14.14^{+1.502}_{-1.513}$ & $13.69^{+0.09}_{-0.09}$ \\

	\hline
	\end{tabular}
	\end{table}

\section{Results and discussion}\label{section:5}
\begin{figure}
	\centering
	\includegraphics[width=10.0cm, height=8.0cm]{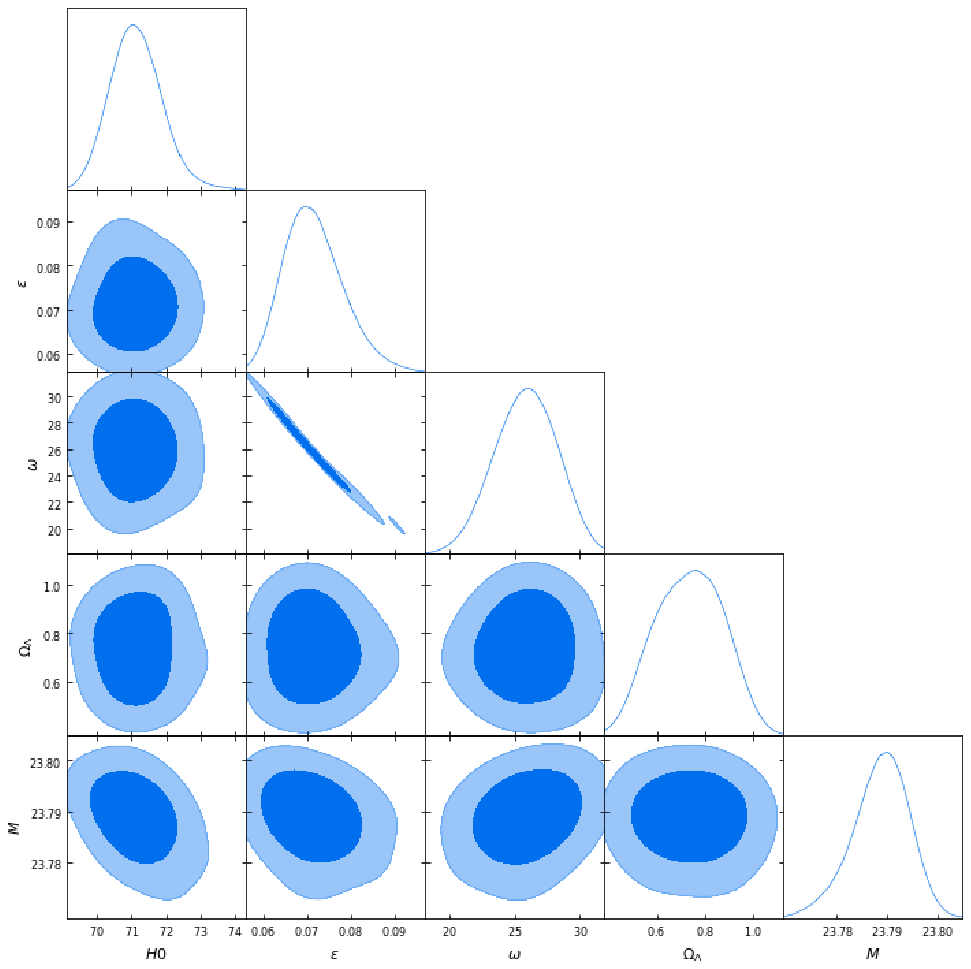}
	\caption{
		One dimensional and two-dimensional marginalized confidence regions for the model parameters of $\Lambda_{H1}$ from DS1 data set }\label{fig:1}
\end{figure}
\begin{figure}
	\centering
	\includegraphics[width=10.0cm, height=8.0cm]{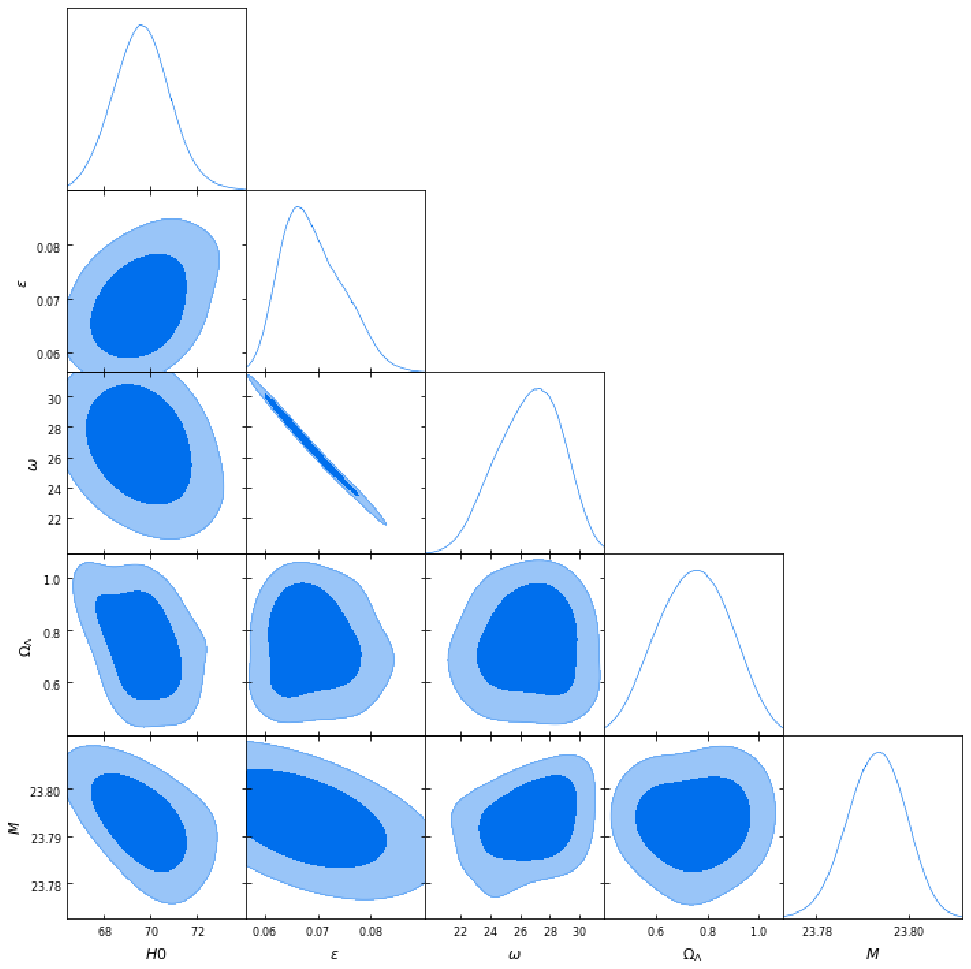}
	\caption{
		One dimensional and two-dimensional marginalized confidence regions for the model parameters of $\Lambda_{H1}$ from DS2 data set}\label{fig:2}
\end{figure}
\begin{figure}
	\centering
	\includegraphics[width=10.0cm, height=8.0cm]{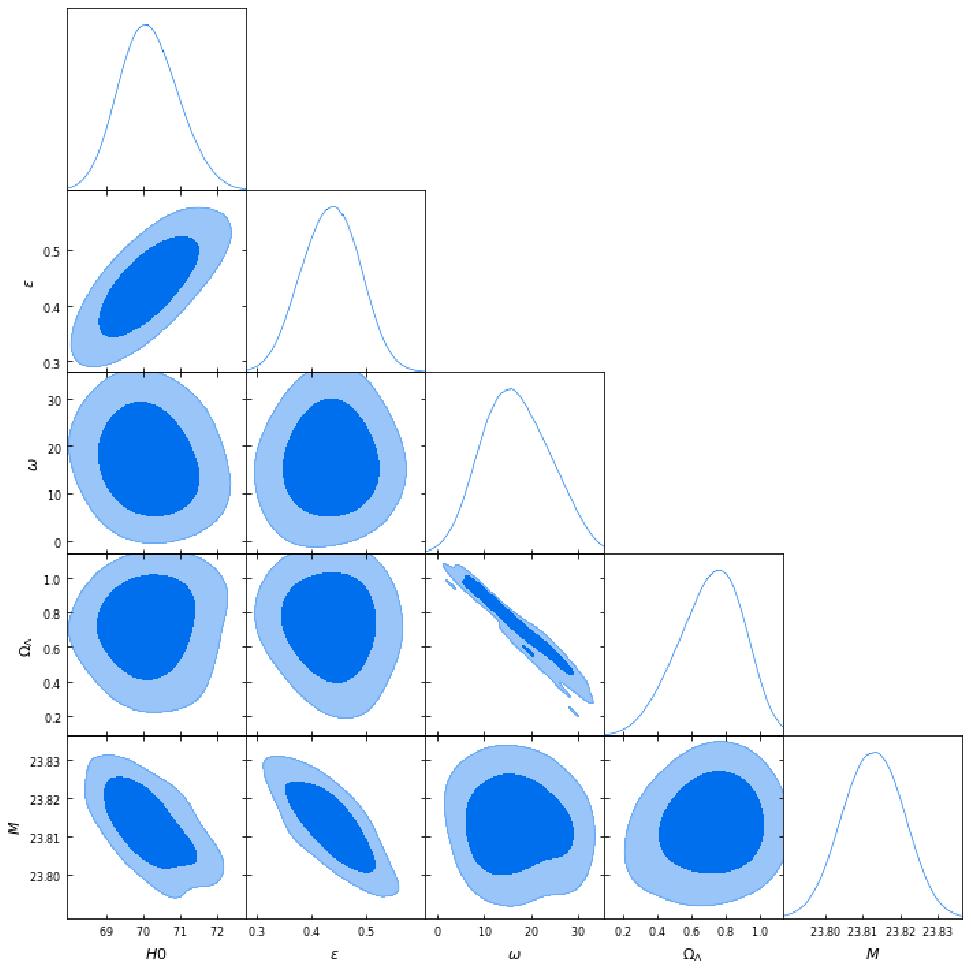}
	\caption{
		One dimensional and two-dimensional marginalized confidence regions for the model parameters of $\Lambda_{H2}$ from DS1 data set}\label{fig:3}
\end{figure}
\begin{figure}
	\centering
	\includegraphics[width=10.0cm, height=8.0cm]{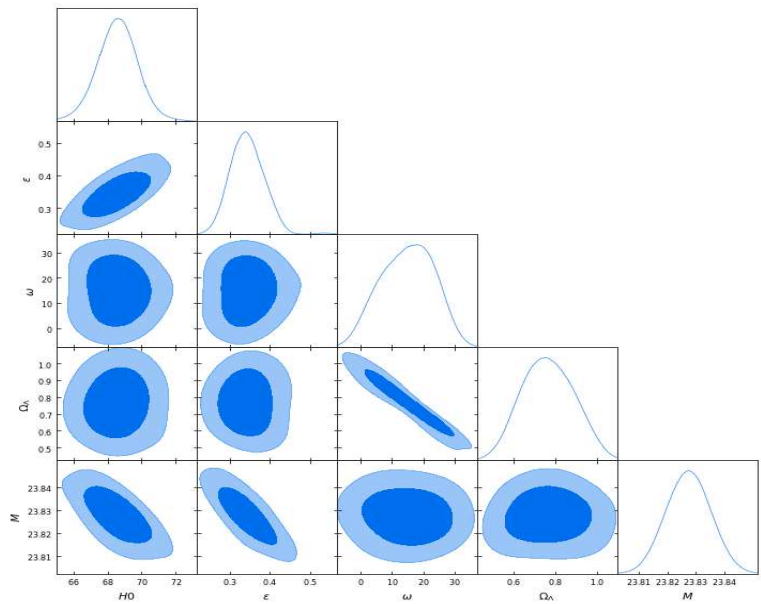}
	\caption{
		One dimensional and two-dimensional marginalized confidence regions for the model parameters of $\Lambda_{H2}$ from DS2 data set}\label{fig:4}
\end{figure}
\jt{In this section, we report the fitting results of the $\Lambda_{H1}$, $\Lambda_{H2}$ and $\Lambda$CDM models using the two data sets  DS1 and DS2 defined in the previous section, and discuss the implications of these results. Figures 1-4 show the constrained parameter space  for the  $\Lambda_{H1}$ and $\Lambda_{H2}$ models under consideration  at $68.3\%$ and $95.4\%$ confidence level (CL) using DS1 and DS2 data sets, respectively. The mean fitting results obtained for both models with $\Lambda$CDM using DS1 and DS2 joint analysis are summarized in Table 1. We report uncertainties corresponding to $1\sigma$ CL. The transition redshift $z_{tr}$, the present values of deceleration parameter $q_0$, effective EoS parameter $w_{eff}(z=0)$, and the present age of the Universe, $t_0$, for these models are given in Table 2. The $\chi^2$, reduced $\chi^2_{red}$ ($=\chi^2/(N-d)$, where $N$ is the number of observational data and $d$ is the number of free parameters), the model selection criterion (AIC, BIC, $\Delta$AIC and $\Delta$BIC) of different models are listed in Table 3. It is to be noted that we have taken $N=83$ for DS1 and $N=82$ for DS2( 40 bin data points of Pantheon, 36 data points of H(z), 06 of BAO/CMB and 01 of $H_0$) and $d=4$ for our joint observational analysis: \jt{($H_0, \epsilon,\omega, \Omega_\Lambda)$}. In what follows, we present the analysis of data in two parts: the cosmological parameters and the model selection criterion.}\\
\begin{figure}[b]
	\centering
	\includegraphics[width=7.5cm, height=7.0cm]{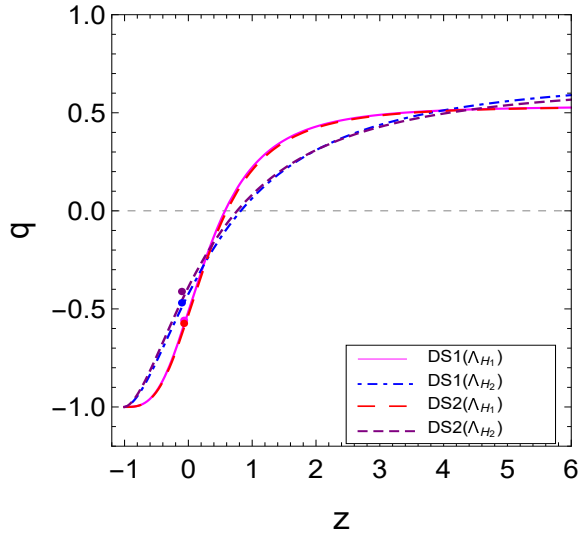}
	\caption{
		The redshift evolution of the deceleration parameter for $\Lambda_{H1}$ and $\Lambda_{H2}$ models obtained from observational data sets DS1 and DS2. A dot denotes the current value of $q$ (hence $q_0$).}\label{fig:5}
\end{figure}
\subsection{Cosmological parameters}
\noindent \jt{The evolution of the deceleration parameter, $q$, with the redshift for the best-fit values of the parameters is shown in Fig. 5. It is observed that there is a sign change in each trajectory of $q(z)$ from positive to negative showing that the Universe transits from decelerated phase to accelerated phase (positive values of $q$ indicate decelerating expansion while negative values indicate an accelerating evolution). We find that the $\Lambda_{H1}$-model transits at around $z_{tr}=0.574^{+0.360}_{-0.324}$ with DS1 data and $z_{tr}=0.601^{+0.339}_{-0.343}$ with DS2 data, which are little smaller than $\Lambda$CDM model. However, model $\Lambda_{H2}$ transits at around $z_{tr}=0.821^{+0.201}_{0.103}$ with DS1 data and $0.763^{+0.245}_{-0.167}$ with DS2 data, which are higher than the transition value of $\Lambda$CDM model. The present-day values of $q_0$ and the transition redshift $z_{tr}$ are listed in Table 2. It is found that the present value of $q$ for $\Lambda_{H1}$ is $q_0=-0.580^{+0.210}_{-0.160}$ using DS1 data and $q_0=-0.610^{+0.176}_{-0.190}$ using DS2 data. However, the present values of $q$ for $\Lambda_{H2}$ model are $q_0=-0.421^{+0.386}_{-0.288}$ and $-0.388^{+0.294}_{-0.630}$ using DS1 and DS2 data, respectively. We observe that both the values of $q_0$ for data set DS1 and DS2 in $\Lambda_{H1}$ model are very close to observational constraint $q_0\simeq -0.63 \pm 0.12 $ \cite{ag1,ag2}, and are smaller than these values in $\Lambda_{H2}$ model. It is to be noted that $q$ tends to $-1$ in late times for both the models.\\
\indent The evolution of the effective EoS parameter $w_{eff}$ with redshift $z$ is plotted in Fig.6 for the different models with their respective best fit values. The present values of $\omega_{eff}$ are listed in Table 2 for data sets DS1 and DS2. The present values of $w_{eff}$ for $\Lambda_{H1}$ model are $w_{eff}(z=0)=-0.675^{+0.173}_{-0.123}$  and $w_{eff}(z=0)=-0.680^{+0.172}_{-0.120}$  whereas for $\Lambda_{H2}$ model, we have $w_{eff}(z=0)=-0.615^{+0.257}_{-0.259}$  and $w_{eff}(z=0)=-0.592^{+0.196}_{-0.435}$ for the data sets DS1 and DS2, respectively. These values are comparatively higher than  the $\Lambda$CDM model. It is also observed from Fig.6 that $w_{eff}$ becomes positive at high redshifts, which represents the early decelerated phase. In late times, $w_{eff}$  approaches to $-1$ for all these models, thus leading to Einstein-de-Sitter behavior. These models do not cross the phantom-divide line $w_{eff}=-1$, which shows that they are free from big-rip singularity. Thus, $w_{eff}$ can easily accommodates both phases of the cosmic evolution, i.e., early decelerated phase and late-time accelerated phase.} \\
\begin{figure}
	\centering
	\includegraphics[width=7.5cm, height=6.5cm]{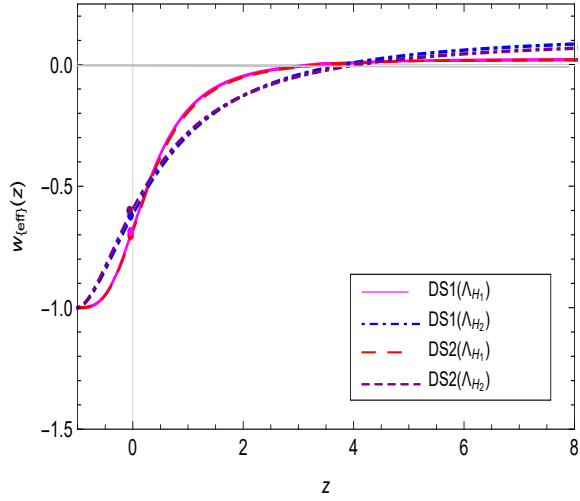}
	\caption{
		The redshift evolution of effective EoS parameter for $\Lambda_{H1}$ and $\Lambda_{H2}$ models using DS1 and DS2 data sets. A dot denotes the present value of the EoS parameter}\label{fig:6}
\end{figure}
\begin{figure}
	\centering
	\includegraphics[width=7.5cm, height=6.5cm]{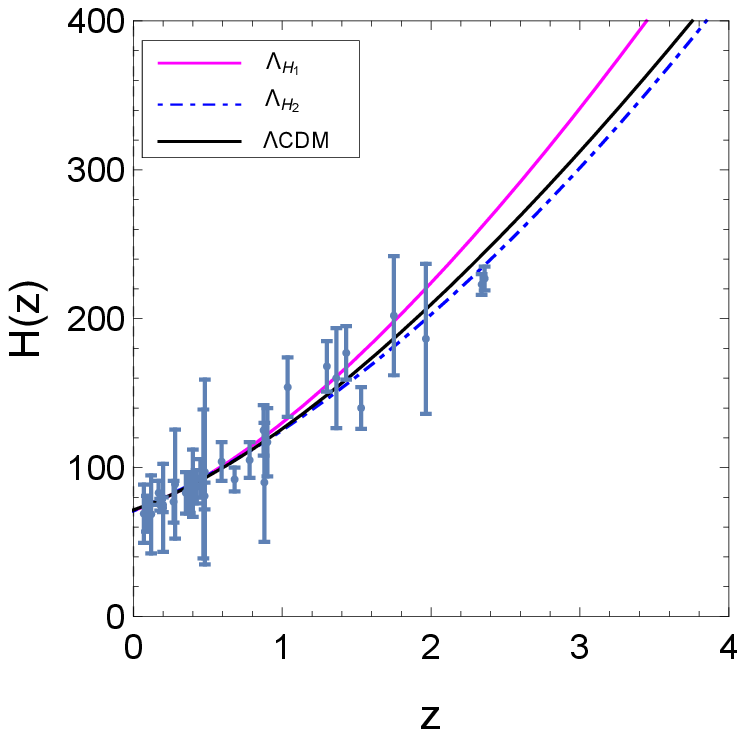}
	\caption{
		Variation of the Hubble function as a function of the redshift $z$ for the best-fit value of the models using DS1 data set. The observational 36 $H(z)$ points are shown with error bars (grey colour).  The variation of the Hubble function in the standard $\Lambda$CDM model is also represented as the solid curve}\label{fig:7}
\end{figure}
\begin{figure}
	\centering
	\includegraphics[width=7.5cm, height=6.5cm]{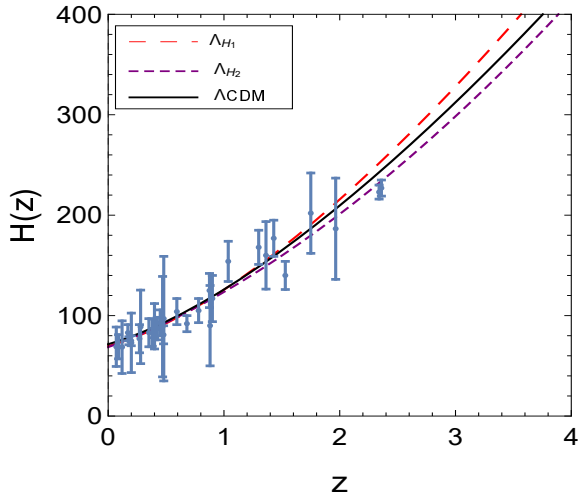}
	\caption{
		Variation of the Hubble function as a function of the redshift $z$ for the best-fit value of the models using Ds2 data set. The observational 36 $H(z)$ points are shown with error bars (grey colour).  The variation of the Hubble function in the standard $\Lambda$CDM model is also represented as the solid curve}\label{fig:8}
\end{figure}
\indent \jt{ The evolutions of the age of the Universe with redshift for the best estimates of model parameters using DS1 and DS2 data sets are given in Table 2.  The age of the Universe for $\Lambda_{H1}$ and $\Lambda_{H2}$ models are
$t_0=13.69^{+1.829}_{-1.548}$ Gyrs  and  $t_0=14.18^{+1.419}_{-1.421}$ Gyrs for DS1 data set, and  $t_0=13.73^{+1.941}_{-1.419}$ Gyrs and $t_0=14.14^{+1.502}_{-1.513}$ Gyrs for DS2 data set, respectively. In our finding, the large errors on the ages of the Universe for models are reported.}\\
\indent \jt{From Table 1, we observe that  the current Hubble constant $H_0$ for $\Lambda_{H1}$, $\Lambda_{H2}$ and $\Lambda$CDM models constrained from DS2 are a bit higher than the constrained observed from the Planck mission ($H_0=67.8\pm0.9$ $Km\;s^{-1}\;Mpc^{-1}$)\cite{h1}. However, the addition of local $H_0$, namely DS1 data set, makes the constraint on $H_0$ bigger for these models. It is to be noted that the improved local measurement $H_0= 73.5 \pm 1.4$ $km \;s^{-1} Mpc^{-1}$ reported by SH0ES \cite{reid} exhibits a strong tension with the Planck mission data \cite{h2}.  \jt{Let us note that in the paper \cite{hei} it has been shown that true quintessence models show a preference for  lower values of  $H_0$ relative to the $\Lambda$CDM model; and this is true even for  coupled quintessence,  as recently shown in  \cite{Adria}.  Let us, however, emphasize that there are dynamical DE models in the market which show an effective quintessence behavior (i.e. they mimic quintessence in that the DE density diminishes with the expansion) but they are nevertheless very different from true quintessence models based on scalar fields. Some of these models mimicking quintessence behavior can have an impact on the $H_0$ tension (and even  on the $\sigma_8$) one. Such is the case e.g.  for the running vacuum models (RVMs), see the recent work \cite{EPLPersp}.   It is also interesting to remark that BD cosmology can help to relax these tensions, as shown in \,\cite{BD1920,js1}, the reason being that BD cosmology with a CC term can mimic the RVM behavior\,\cite{GRF2018}. Our present study further reinforces such welcome property of BD models possessing vacuum energy}. \\
\indent Figures 7 and 8 display the Hubble diagram with the error bar of Hubble data set in the range $z\in (0, 4)$ for DS1 and DS2 data sets. For the sake of comparison, the  flat $\Lambda$CDM scenario is also shown. The evolutions of $H(z)$ of $\Lambda_{H1}$ and $\Lambda_{H2}$ model are comparatively similar to the $\Lambda$CDM model.  At low redshifts, the cosmological evolutions of  models $\Lambda_{H1}$ and $\Lambda_{H2}$ are consistent with the Hubble data.} \\
\begin{table}[h]
	\caption{Summary of $\chi^2$, $\chi^2_{red}$, $AIC$ and $BIC$ values and their differences from the reference model of $\Lambda CDM$ obtained from joint analysis of DS1 and DS2 data sets}
	\centering
	\begin{tabular}{c | c c c c c c c}
		\hline \hline
		Data & Model & $\chi^2$ & $\chi^{2}_{red}$ & $AIC$ & $BIC$ & $\Delta AIC$ & $\Delta BIC$ \\
[0.5ex]
		\hline\\
DS1 & $\Lambda$CDM & $34.78$ & $0.43$ & $40.78$ & $48.04$ & $0$ & $0$\\\\
$$	&	 $\Lambda_{H1}$ & $34.94$ & $0.44$ & $42.94$  &  $52.62 $ & $2.16$ & $4.58$\\\\
$$ &		$\Lambda_{H2}$ & $36.94$ & $0.47$ & $44.94$ &  $54.62$ & $4.16$ & $6.58$\\\\
\hline\\
DS2 & $\Lambda$CDM & $34.69$ & $0.44$ & $40.69$ & $47.91$ & $0$ & $0$ \\\\
 $$ &     $\Lambda_{H1}$ & $35.28$ & $0.45$ & $43.28$ & $52.90$ & $2.59$ & $4.99$\\\\
  $$ &    $\Lambda_{H2}$ & $39.42$ & $0.51$ & $47.42$ & $57.04$ & $6.73$ & $9.13$\\

		\hline
	\end{tabular}
	\label{table:nonlin}
\end{table}
\subsection{Model selection}
\noindent \jt{Taking into account that models  $\Lambda_{H1}$ and $\Lambda_{H2}$  have the same number of extra  parameters ($\lambda$ and $\sigma$, respectively), we could directly compare them on the basis of computing the minimum $\chi^2$ values for each model. But when we compare them with  the concordance $\Lambda$CDM model, the $\chi^2$ comparison becomes unfair because in the context of BD theory the number of parameters is different. In our approach  two  more parameters, $\epsilon$ and $\omega$, have to be considered  which are not involved in the concordance model.  For this reason we employ the Akaike Information Criterion (AIC)\cite{ak} and the Bayesian Information Criterion (BIC)\cite{bs} so as to do a fairer model comparison, see e.g.  \cite{liddle} for a review}.  The AIC parameter is defined through the relation
\begin{equation}\label{b9}
AIC=\chi^{2}_{min}+2d,
\end{equation}
where $d$ is the number of free parameters in the model and $\chi^{2}_{min}$ is the minimum value of the $\chi^{2}$ function. We calculate $\Delta AIC_{i}=AIC_{i}-AIC_{j}$, where $i,j$ denote respectively the model $i$ and model $j$. This is interpreted as ``evidence in favour" of the model $i$ compared to the model $j$. The preferred  model for this criterion is one with the smaller value of AIC.\\
\indent On the other hand, the BIC is defined through the relation
\begin{equation}\label{b10}
BIC=\chi^2_{min}+d\;ln\;N,
\end{equation}
where $N$ is the number of data points. Similar to $\Delta AIC$, we have $\Delta BIC= BIC_{i}-BIC_{j}$. This can be interpreted as ``evidence against" the model $i$ compared to the model $j$.\\
\indent  To be more precise, a model having $0 \le \Delta AIC <2$ and $0 \le \Delta BIC <2$  receives  ``strong evidence in favour". In contrast,  for $2 < \Delta AIC <4$ and for $2 \le \Delta BIC <6$, the model has ``average evidence in favour", whereas for $4 < \Delta AIC <7$ and $6 \le \Delta BIC <10$, the model is considered to have  ``less evidence in favour"; and, finally,  for $\Delta AIC>10 $ or $\Delta BIC >10$, the model receives no significant support since it has  ``no evidence in favour" \cite{liddle}.\\
\indent In Table 3, we present the values of $\chi^2$, AIC, BIC and their differences for the discussed models. From Table 3, we find that $\Lambda_{H1}$ has $\Delta$AIC$\sim 2.16$ and $\Delta$AIC$\sim 2.59$, whereas it has $\Delta$BIC$\sim 4.58$ and $\Delta$BIC$\sim 4.99$ from DS1 and DS2 data sets, respectively. We see that the $\Lambda_{H1}$ model is in the range of $2 \le \Delta AIC <4$ and $2 \le \Delta BIC <6$. Thus, this model shows average evidence in favour. However, the $\Lambda_{H2}$ model has $\Delta$AIC$\sim 4.16$ and $\Delta$AIC$\sim 6.73$, and $\Delta$BIC$\sim 6.58$ and $\Delta$BIC$\sim 9.13$ from DS1 and DS2 data sets, respectively. Since, this model shows the differences in the range of $4 < \Delta AIC <7$ and $6 \le \Delta BIC <10$, therefore, this model has ``less evidence in favour". The AIC and BIC impose a strict penalty against the presence of additional parameters.\\
\indent \jt{In Table 3, we give the values of $\chi^2_{min}$. We find, among the models, the $\Lambda$CDM model is still the best one in fitting the current observational data. The $\Lambda$CDM has least number of parameters, but it gets the smallest $\chi^2_{min}$ value in this fit. The $\Lambda_{H1}$ and $\Lambda_{H2}$ models have one more parameter than the $\Lambda$CDM  model. However, $\Lambda_{H1}$ yields  very close $\chi^2_{min}$ values to that of $\Lambda$CDM from the two datasets ,DS1 and DS2, whilst $\Lambda_{H2}$ renders  corresponding  higher values. The reduced $\chi^2_{red}$ of both the models is less than one (cf. Table 3),  so overall these two models can be considered in good agreement with  the $\Lambda$CDM model and data are consistent with the considered models. Model $\Lambda_{H1}$ seems to be the closest. Therefore, our analysis suggests that the BD version of the $\Lambda$-cosmology is on an essentially equal footing position as compared to the concordance model in the light of our fits.}
\section{Conclusion}\label{section:6}
\label{conclusion}
\jt{Among the many proposals to describe the late time acceleration of the Universe, the cosmological constant (CC) is  the simplest candidate to provide an explanation.  It defines the  standard or concordance $\Lambda$CDM model,  and it is referred to also as  the $\Lambda$-cosmology.  In this paper, we have studied if the $\Lambda$-cosmology, which in its standard version is implemented through general relativity (GR), can be realized too  in the context of Brans-Dicke (BD) gravity and with a similar or better level of achievement.  We have assumed, as in many other studies in the literature, that the local constraints imposed on  BD gravity can be avoided by resorting to the presence of screening forces, which do not affect the study of cosmology at the level of the large scales. To make our study of BD cosmology more complete, we have explored the possibility to add some dynamical component to the vacuum energy in the BD framework.  For such purpose we have explored theoretical and observational features of a simple class of cosmological models driven by a time-varying vacuum energy density for a spatially flat FRW spacetime in BD theory, hence beyond the GR paradigm underlying the  standard $\Lambda$CDM model.  Such class of models is characterized by a time-evolving CC of the form  $\Lambda=\lambda+\sigma H$.  We have solved these models searching for power-law solutions and checked the consistency of the obtained solutions. We have  separately solved the two particular cases  $\Lambda=\lambda$ (model $\Lambda_{H1}$) and $\Lambda=\sigma H$ (model $\Lambda_{H2}$) as well as the general case with arbitrary $\lambda$ and $\sigma$.  For the numerical analysis we have used the latest observational measurements of SnIa (Pantheon), $H(z)$, BAO/CMB and local $H_0$. The corresponding results are presented in Table 1 while Figs. 1-4 show the confidence contours for the different parameters.  The two models  $\Lambda_{H1}$ and  $\Lambda_{H2}$ have been analyzed using two data sets DS1 and DS2, where  the local $H_0$ value is only included in the first set.  Upon using the  model selection criteria AIC and BIC, we find that for both data sets  the phenomenological performance of model $\Lambda_{H1}$ is better than that of  $\Lambda_{H2}$, but both models are acceptable for the description of the data.  The more general model   $\Lambda=\lambda+\sigma H$ only interpolates between the two former ones.  Since the description of the data does not improve with a nonvanishing value of $\sigma$, we conclude that the BD version of the  $\Lambda$-cosmology is the preferred option and it proves comparable to the conventional $\Lambda$CDM  in light of our fitting results.}\\
\indent \jt{The observational analysis shows that both the models $\Lambda_{H1}$ and $\Lambda_{H2}$ exhibit the same characteristics concerning the evolution of the Universe, i.e., they describe the transition  from a decelerated to an accelerated phase.  In both cases the effective equation of state (EoS) performs an evolution from $w_{eff}=0$ in the remote past to to $w_{eff}=-1$ in the remote future without crossing the phantom divide $w=-1$.  At late-time, the deceleration parameter $q$ tends also to $-1$,  showing that these models predict de Sitter behavior in the future. In the case of  $\Lambda_{H1}$, the current value is  $q_0\simeq -0.6$, thus similarly to the concordance model, whereas for model $\Lambda_{H2}$ it is smaller in absolute value ( $q_0\simeq -0.4$), see Table 2 for detailed results.} \\
\indent Using the fitting values of the parameters listed in Table 1 into  the consistency relation  \eqref{consistency1}, it has been found that the $\Lambda_{H1}$ model has $\Omega_{\Lambda}=0.728^{+0.213}_{-0.173}$ with DS1 data set and $\Omega_{\Lambda}=0.733^{+0.222}_{-0.146}$ with DS2 data set. The errors are sizeable, but even at the level of the best fit values these results are perfectly consistent with those  obtained from the observations using DS1 and DS2 data sets, viz., $\Omega_{\Lambda}=0.73^{+0.133}_{-0.183}$ and $\Omega_{\Lambda}=0.72^{+0.142}_{-0.169}$, respectively (cf. Table 1).  As for model $\Lambda_{H2}$, using the fitting values of that table into the consistency relation \eqref{consistency2} we find $\Omega_{\Lambda}=-1.658^{+0.826}_{-11.361}$ with DS1 data and $\Omega_{\Lambda}=-2.025^{+0.985}_{-2.722}$ with DS2 data set. These numerical values are inconsistent with the fitting values of $\Omega_{\Lambda}$  obtained for DS1 and DS2. The latter remain nonetheless in the approximate  range  $\Omega_{\Lambda}=0.72-0.77$ (using the errors) for both data sets (cf. Table 1).  Therefore, we find  that in the context of our analysis  the performance of model $\Lambda_{H1}$  is very similar to that of the standard $\Lambda$CDM model. At the level of information criteria it is at the border line of not implying any significant difference with the standard $\Lambda$-cosmology. On the other hand, $\Lambda_{H1}$ satisfies remarkably well the consistency equation \eqref{consistency1}.  In stark contrast, despite the quality fit of the  $\Lambda_{H2}$  is lesser, it is still a reasonable one for the DS1 data while it is not so good for DS2. In addition, the model does not adapt to the consistency condition \eqref{consistency2}.  We should emphasize that this is not caused by any  analytical inconsistency in our study, the discrepancy  is only numerical because of assuming a power-law relation between the scalar field and the scale factor. Such relation may not be a perfect choice for  the solutions of the BD field equations in the case of the $\Lambda_{H2}$ model, and this means that a more general family of solutions is needed. Let us, however, note that it is difficult to explore other kind of analytical solutions for the complicated system of BD equations and one may be forced to go fully numerical in this case.  The dynamics of model $\Lambda_{H1}$, instead, adapts  well to the power-law solution since the numerical consistency is manifest.  Recall from the footnote on page 9 that model  $\Lambda_{H1}$ has a smooth analytic limit to GR for $\epsilon\to 0$.  In the case of model  $\Lambda_{H2}$  we do not expect such limit to hold since its effective cosmological term is not constant at any time and hence there is no smooth connection with the concordance $\Lambda$CDM model.  This can also explain why this model does not adapt equally well to the same power-law family of solutions as for model  $\Lambda_{H1}$. \\
\indent The main conclusion  of our study is in our opinion significant. We have shown that model $\Lambda_{H1}$, namely the $\Lambda$-cosmology in the context of the BD theory, is more favored than $\Lambda_{H2}$ and is comparable to the concordance $\Lambda$CDM  model within GR. Since for model $\Lambda_{H1}$  the consistency relation is fully realized also at the numerical level we can say that this claim is robust. In the case of model $\Lambda_{H2}$ it is only indicative. This does not preclude, however, the possibility that other forms of dynamical $\Lambda$  can improve the performance of BD theory as compared to the standard  $\Lambda$CDM model. In the meantime our analysis shows that of all the possible dynamical models $\Lambda=\lambda+\sigma H$ within the BD paradigm, the most promising ones are those with $\sigma\simeq 0$. Notwithstanding, this conclusion should not be interpreted as saying that a rigid cosmological term in BD theory is equivalent to the effect of a rigid cosmological term in GR.  As previously noted, model $\Lambda_{H1}$ despite it being associated to a rigid cosmological constant term in the BD context, it is perceived as a running vacuum model from the point of view of GR. This fact is helpful since it is known that the running vacuum model performs a fit to the overall cosmological data which is competitive with that of the concordance model\,\cite{EPLPersp}.  Only future studies can reveal if Brans-Dicke gravity with a rigid cosmological term can be fully competitive with GR in all aspects of the observational cosmological data, and to which extent it may be necessary to introduce a dynamical component in it.  Here we have shown that the $\Lambda$-cosmology in such BD context is not second rate as compared to the GR version, and that the addition of the simplest possible  dynamical component to the vacuum energy  does not perturb exceedingly this conclusion.   \\

\begin{acknowledgements}
One of the authors, JSP, acknowledges partial support by  projects  PID2019-105614GB-C21 and FPA2016-76005-C2-1-P (MINECO, Spain), 2017-SGR-929 (Generalitat de Catalunya) and CEX2019-000918-M (ICCUB). JSP also acknowledges participation in the COST Association Action CA18108 ``{\it Quantum Gravity Phenomenology in the Multimessenger Approach (QG-MM)}''.

\end{acknowledgements}

\section*{\jt{Appendix A: Time-varying model  $\Lambda=\lambda+\sigma H$  within the BD theory}}
\noindent In this appendix, we briefly provide the analytical solution of  the  time varying $\Lambda(t)$ model
 $\Lambda=\lambda+\sigma H$, as defined in Eq.\eqref{c12} of the main text, in its general form, i.e.  for arbitrary values of $\lambda \neq 0$ and $\sigma \neq 0$, and within the Brans-Dicke context.\\
\indent Using this form of $\Lambda(t)$ into Eq. \eqref{c11}, the evolution equation for the Hubble function can be rewritten as
\begin{equation}\label{c41}
\dot{H}+\frac{(3+\epsilon)}{2}H^2-\frac{3\sigma}{(6+6\epsilon-\omega \epsilon^2)}H-\frac{3\lambda}{(6+6\epsilon-\omega \epsilon^2)}=0.
\end{equation}
The solution to this differential equation reads as follows:
\begin{equation}\label{c42}
H=\jt{H_0}\left[\frac{3\sigma}{(6+6\epsilon-\omega \epsilon^2)(3+\epsilon)}+\alpha\left(\frac{e^{(3+\epsilon)\alpha\; t}+1}{e^{(3+\epsilon)\alpha\; t}-1}\right)\right],
\end{equation}
where $\alpha=\frac{\sqrt{9\sigma^2+6\lambda(6+6\epsilon-\omega \epsilon^2)(3+\epsilon)}}{(6+6\epsilon-\omega \epsilon^2)(3+\epsilon)}$.\\
\noindent The solution of the scale factor is given by
\begin{equation}\label{c43}
a(t)=\left(e^{(3+\epsilon)\alpha\; t}-1\right)^{\frac{1}{3+\epsilon}}\;e^{\left(\frac{3}{(6+6\epsilon-\omega \epsilon^2)(3+\epsilon)}-1\right)\alpha\; t}.
\end{equation}
In this case, the deceleration parameter is obtained as
\begin{equation}\label{c44}
q=-1+\frac{2\alpha^2(3+\epsilon)\;e^{(3+\epsilon)\alpha\;t}}{\left(\frac{3\sigma}{(6+6\epsilon-\omega \epsilon^2)(3+\epsilon)}+\alpha\left(\frac{e^{(3+\epsilon)\alpha\; t}+1}{e^{(3+\epsilon)\alpha\; t}-1}\right)\right)^{2}\;
\left(e^{(3+\epsilon)\alpha\; t}-1\right)^{2}}.
\end{equation}
Lastly, the effective EoS parameter can be worked out with the following result:
\begin{equation}\label{c45}
w_{ef}=-1+\frac{2}{3}\frac{2\alpha^2(3+\epsilon)\;e^{(3+\epsilon)\alpha\;t}}{\left(\frac{3\sigma}{(6+6\epsilon-\omega \epsilon^2)(3+\epsilon)}+\alpha\left(\frac{e^{(3+\epsilon)\alpha\; t}+1}{e^{(3+\epsilon)\alpha\; t}-1}\right)\right)^{2}\;
\left(e^{(3+\epsilon)\alpha\; t}-1\right)^{2}}.
\end{equation}
The above solution of cosmological parameters are found in terms of exponential form which show that the model can accommodate the late time acceleration. It can also be shown that $\Lambda_{H1}$ and $\Lambda_{H2}$ models are particular solutions of the above general vacuum model.

\end{document}